\documentclass[useAMS,usenatbib]{mn2e}
\usepackage{graphicx}
\newif\ifAMStwofonts

\newcommand{\msun}   {{\rm M}_\odot} 
\newcommand{\Msun}   {$\msun$}

\newcommand{\HI}     {{\sc H i}}
\newcommand{\HII}    {{\sc H ii}}

\newcommand{\oversim}[2]{\protect{\mbox{\lower0.5ex\vbox{%
  \baselineskip=0pt\lineskip=0.2ex
  \ialign{$\mathsurround=0pt #1\hfil##\hfil$\crcr#2\crcr\sim\crcr}}}}}
\newcommand{\simless} {\mbox{$\,\mathrel{\mathpalette\oversim<}\,$}} % <~ sign

\title[Chemical evolution with variable IMFs]{A Possible Origin of the 
      Mass--Metallicity Relation of Galaxies}
\author[J. K\"oppen, C. Weidner and P. Kroupa]
       {J. K\"oppen$^{1, 2,
           3}$\thanks{e-mail: koppen@astro.u-strasbg.fr},
         C. Weidner$^{4, 5}$\thanks{e-mail: cweidner@astro.puc.cl}
         and P. Kroupa$^{5,
           6}$\thanks{e-mail: pavel@astro.uni-bonn.de}\\ 
        $^1$ UMR 7550, Observatoire Astronomique,
             11 rue de l'Universit\'e, F-67000 Strasbourg, France \\
        $^2$ Institut f\"ur Astronomie und Astrophysik
             der Universit\"at Kiel, D-24098 Kiel, Germany \\
        $^3$ International Space University, Parc d'Innovation,
             F-67400 Illkirch, France \\
	$^4$ Departemento de Astronom{\'i}a y Astrof{\'i}sica, Pontificia
      Universidad Cat{\'o}lica de Chile, Av. Vicu{\~n}a MacKenna 4860,\\
      Macul, Santiago, Chile\\  
        $^5$ Argelander Institut f\"ur Astronomie der Universit\"at
        Bonn, Auf dem H\"ugel 71, D-53121 Bonn, Germany \\
        $^6$ The Rhine-Stellar-Dynamical Network
        }

\date{Accepted \today .
      Received \today ;
      in original form \today}

\pagerange{\pageref{firstpage}--\pageref{lastpage}}
\pubyear{2006}

\begin{document}
\bibliographystyle{mn2e}
  \maketitle

  \label{firstpage}

  \begin{abstract}
Observations show that galaxies follow a mass--metallicity relation
over a wide range of masses. One currently favoured explanation is
that less massive galaxies are less able to retain the gas and stellar 
ejecta and thus may lose the freshly produced metals in the form of 
galactic outflows. Galaxies with a low current star formation rate
have been found to contain star clusters up to a lower mass limit. Since
stars are predominately born in clusters, and less massive clusters 
have been found to be less likely to contain very massive stars,
this implies that in environments or at times of low star formation,
the stellar initial mass function does not extend to as high masses
as during high star formation epochs. It is found that the oxygen
yield is reduced by a factor of thirty when the star formation rate is
decreased by 3 to 4 orders of magnitude. With this concept, chemical
evolution models for galaxies of a range of masses are computed and
shown to provide an excellent fit to the mass--metallicity relation
derived recently by \citet{THK04}. Furthermore, the models match the
relation between galaxy mass and effective yield. Thus, the scenario
of a variable integrated stellar initial mass function, which is based
on the concept of formation of stars in clusters, may offer an
attractive alternative or partial explanation of the mass--metallicity
relation in galaxies.
  \end{abstract}


  \section{Introduction}

  It has been known for a long time that there exist fairly well-defined
  relations between the metallicity and mass (or luminosity) among gas-rich 
  galaxies \citep{LRS79,SKH89} as well as among gas-poor galaxies
  \citep{Fa73,BH91}. \citet{ZKH94} point out the similarity of these
  relations, which is suggestive of a common physical origin of this
  phenomenon. Recently, the wealth of data from the Sloan Digital Sky Survey
  permitted \citet{THK04} to study the mass--metallicity relation over a wide
  range of masses and metallicities for about 53000 star-forming galaxies.

  \citet{THK04} interpret their data with a scenario of gas outflows which 
  are stronger in dwarf galaxies. More massive galaxies are 
  able to retain the gas longer and more effectively than low mass
  objects, thus the metallicity can build up to higher values because
  following generations of stars will be formed in an enriched gas
  environment, while the low-mass objects lose their metals through
  galactic winds. Different amounts of dark matter can also assist in
  this process \citep{DS86}. \citet{DKW04} show that the chemical evolution
  in hierarchical galaxy formation models are able to explain the observed
  mass--metallicity relation and other relations like the Tully-Fisher
  relation. This is achieved by including feedback processes like galactic
  winds into their cosmological simulations. The formulation of the feedback
  includes free parameters such as the feedback efficiency or the yield.

  While galactic outflows are observed and this scenario provides a
  satisfactory match of the observed relation, one should not yet
  exclude other explanations or contributions from other
  processes. This is particularly important because the current
  galactic wind models still need to be individually-fitted to the
  galaxies and are not easily explained by physical parameters of the
  galaxy. Furthermore, recent works by \citet{LSC06} and \citet{Da06}
  show that simple outflows of ISM gas have difficulties reproducing
  the yields in dwarf galaxies. \citet{LSC06} used observations with {\it
  Spitzer} to derive the mass--metallicity relation from a number of dwarf
  irregular galaxies. They conclude that the dispersion in the
  relation and the large variations in the effective yields ``are
  difficult to understand if galactic superwinds or outflows are
  responsible for low metallicities at low mass or luminosity.'' In
  \citet{Da06} a series of closed-box chemical evolution models
  including infall and outflow is presented. There it is shown that
  neither simple infall nor outflow can reproduce the observed low
  effective yields in low mass galaxies, but metal-enriched outflows
  can do. That is, effectively the freshly synthesised elements need
  to be removed from the matter cycle, which is, in principle,
  equivalent to reducing the number of massive stars.

  In this work we like to explore further the implications of the fact that 
  stars form predominantly in clusters \citep{LL03}. \citet{KW03} showed that 
  the integrated galactic IMF (IGIMF) of all stars in a galaxy depends on the
  embedded cluster mass function (ECMF), in the sense that more massive
  clusters are likely to have more massive stars, because of their greater
  amount of available gas to form stars. Based on the relation found by
  \citet{WKL04} between the maximum mass of clusters in a galaxy and the
  current star formation rate (SFR), \citet{WK05a,WK05b} show that the
  high mass part of the resulting IGIMF is strongly dependent on the
  star formation history (SFH) in a galaxy, being in general steeper
  than the IMF. This gives an important influence on the production
  rate of metals \citep{WK05b}. \citet{GP05} explored the effect of
  the IGIMF on the metal production in galaxies. They found a change
  in the general metal yield by a factor of $\sim$ 1.8 and by a factor
  of $\sim$ 1.5 for oxygen and  magnesium when they use the ECMF slope of
  \citet[][i.e.~$\beta$ = 2]{LL03}.

  One thus expects that the effective metal yields from the average stellar
  population may vary with time and galaxy type.  
  It is not the aim of this contribution to exclude the scenario of galactic 
  outflows, but to explore the possibilities and consequences arising purely
  from clustered star formation, and to probe to what extend it may 
  provide an alternative or supporting explanation.

  Section~\ref{se:varimf} briefly reviews the model of the SFR-dependent
  IGIMF (henceforth IGIMF for simplicity), 
  section~\ref{se:oxy} discusses the importance of the oxygen yields
  for this study, while in section~\ref{se:chevo} the chemical
  evolution model is introduced and the effect of a variable IGIMF on
  the chemical evolution of galaxies is studied. In section~\ref{se:discuss}
  the implications of these results are discussed, while
  section~\ref{se:conclu} presents the conclusions from these results.

\section{the variable imf}
        \label{se:varimf}

  In \citet{KW03} and \citet{WK05a,WK05b} it has been shown that the 
  formation of stars predominantly in clusters has a profound influence 
  on the integrated galactic stellar initial mass function (IGIMF) of 
  all stars in a galaxy. In contrast to the IGIMF, which is the
  ``IMF'' integrated over a whole galaxy, the IMF of stars formed in a
  single cluster is well described by the well-known power-law
  function with the Salpeter/Massey slope $\alpha = 2.35$
  \citep{Sal55,Mass03} for stars more massive than $1\,M_{\odot}$, as
  found in several observational studies
  \citep{MH98,SND00,SND02,PaZa01,Mass02,Mass03,WGH02,BMK03,PBK04},
  with 
  \begin{equation}
      \xi(m) \propto m^{-\alpha_{i}}, 
  \end{equation}
  where $\xi(m)\,dm$ is the number of stars in the mass interval $m,\,m+dm$. 
  Below 0.5 $M_{\odot}$ the IMF is found to flatten
  \citep{MS79,Sc86,KTG93,RGH02}, and the {\it standard} IMF, becomes a
  segmented power-law \citep{Kr01,Kr02}: 
  \begin{equation}
     \begin{array}{l@{\quad\quad,\quad}l}
        \alpha_0 = 0.30&0.01 \le m/{M}_\odot < 0.08,\\
        \alpha_1 = 1.30&0.08 \le m/{M}_\odot < 0.50,\\
        \alpha_2 = 2.35&0.50 \le m/{M}_\odot, \le m_{\rm max*}.\\
     \end{array}
     \label{eq:Kroupa-IMF}
   \end{equation}
  \noindent
  Note that this form constitutes a two-part power-law {\it stellar} IMF
  with the ``Salpeter/Massey'' (or short ``Salpeter'') slope above 0.5
  $M_\odot$. Note also that this form is similar to \citet{Ke83}
  two-part power-law form, except that the slope at the high-mass end
  is less-steep here (being the Salpeter/Massey value), and that the
  break in the power-law occurs at a smaller mass here. These
  differences, while fairly subtle, are important as the form used
  here is based on a very thorough star-count analysis using post-1983
  data \citep{KTG93,Kr01}. The physical stellar upper mass limit is
  assumed to be $m_{\rm max *}$ = 150 $M_{\odot}$, following the
  recent results by \citet{WK04}, \citet{Fi05}, \citet{OC05} and
  \citet{Ko06}.

  The masses of young, embedded star clusters also follow  
  a power-law distribution, the embedded cluster mass function (ECMF): 
  \begin{equation}
      \xi_{\rm ecl}(M_{\rm ecl}) \propto M_{\rm ecl}^{-\beta}, 
  \end{equation}
  where $\xi_{\rm ecl}(M_{\rm ecl})~dM_{\rm ecl}$ is the number of embedded 
  clusters in the mass interval ($M_{\rm ecl}$, $M_{\rm ecl} + dM_{\rm ecl}$), 
  and $M_{\rm ecl}$ is the mass in stars. \citet{LL03} find a slope $\beta
  \approx 2$ in the solar neighbourhood for clusters with masses between 50
  and 1000 $M_{\odot}$, while \citet{HEDM03} find $2 \simless \beta \simless
  2.4$ for  $10^{3} \simless M_{\rm ecl}/M_{\odot} \simless 10^{4}$ in the SMC
  and LMC, \citet{ZF99} find $1.95 \pm 0.03$ for $10^4 \simless M_{\rm
  ecl}/M_\odot \simless 10^6$ in the Antennae galaxies. \citet{WKL04} find
  that $\beta = 2.35$ best reproduces the observed correlation between the
  brightest young cluster and the global star formation rate for a sample of
  late-type galaxies.

  The IMF of all stars ever formed in a galaxy (the IGIMF) is obtained by
  summing up all standard IMFs of all clusters from which a galaxy is formed, 
  \begin{equation}
     \xi_{\rm IGIMF}(m) = \int_{M_{\rm ecl,min}}^{M_{\rm ecl,max}} 
      \hspace*{-0.8cm}
             \xi(m\le m_{\rm max}(M_{\rm ecl}))\, 
              \xi_{\rm ecl}(M_{\rm ecl})\,dM_{\rm ecl}, 
    \label{eq:IGIMF1}
  \end{equation}
  where $\xi_{\rm ecl}(M_{\rm ecl})$ is the ECMF and 
  $\xi(m\le m_{\rm max}(M_{\rm ecl}))$ is the stellar IMF in a particular 
  cluster within which the maximal mass of a star is $m_{\rm max}$. 
  A critical ingredient of this concept is the existence of the correlation 
  between the maximum stellar mass $m_{\rm max}$ and the cluster mass $M_{\rm
  ecl}$, which depends on the sequence with which stars of different masses
  are formed. \citet{WK05b} find that several young clusters are best
  described if one assumes that, starting at the lowest mass, stars of
  progressively higher mass are formed.

\begin{figure}
      \centering
      \includegraphics[width=8cm]{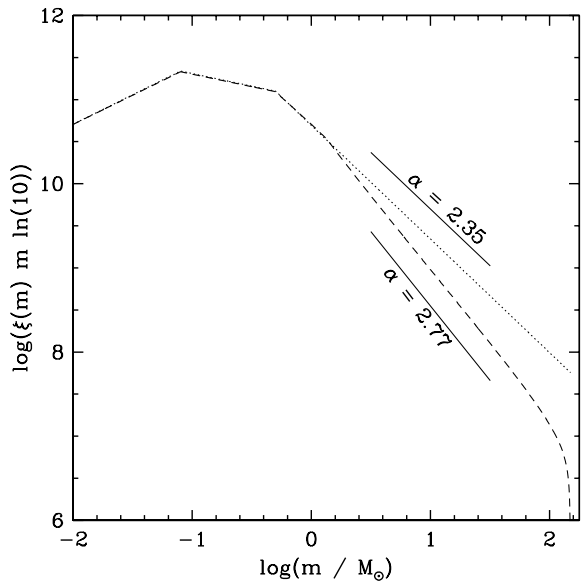} 
\vspace*{-2cm}
      \caption[]{The dotted line is the standard IMF $\xi(m)$, given by
        Eq.~\ref{eq:Kroupa-IMF} which has $\alpha=2.35$ for $m>0.5\,M_\odot$. 
        The dashed line is the overall IGIMF, $\xi_{\rm IGIMF}(m)$, for
        $\beta=2.2$.  
        All functions are scaled to have the same number of objects in the
        mass interval $0.01-1.0\,M_\odot$. Slopes $\alpha=2.35$ and
        2.77 are indicated by lines. Note that we use the notation $\log =
        \log_{10}$ throughout the paper.} 
      \label{f:imfex}  
   \end{figure}

  In Fig.~\ref{f:imfex} we show an example of the resulting IGIMF from
  Eq.~\ref{eq:IGIMF1}, using an ECMF power-law slope $\beta = 2.2$.
  Note that the IGIMF ({\it dashed line}), which has a slope above
  $1.5\,M_{\odot}$ of about 2.77, is steeper than the standard IMF
  ({\it dotted line}); furthermore it turns down rather sharply near
  $m_{\rm max*}=150\,M_\odot$,  the fundamental upper mass limit for
  stars \citep{WK04,WK05a,Fi05,OC05,Ko06}.

  \citet{Em06} puts forward the view that the IGIMF equals the IMF, but we
  note that his discussion points to a few examples which actually do
  impose a variable IGIMF. Our formulation tests, like that of
  Elmegreen, the case of $\beta = 2$, but ours allows a variation of
  the maximum mass of the stars with the SFR, in contrast to
  Elmegreen's where this mass comes out to be constant. 

\subsection{The connection to the star formation rate}
  The IGIMF depends not only on the ECMF but also on the star formation 
  history of a galaxy. \citet{WKL04} derive a relation between the 
  current SFR and the maximum mass of a cluster in a galaxy. With this
  relation 
  \begin{equation}
    \label{eq:Meclsfr}
       \log M_{\rm ecl,max}(t) = \log k_{\rm ML} + 
              0.75 \cdot \log{{\rm SFR}(t)} + 6.77,
  \end{equation}
  and a mass-to-light ratio of $k_{\rm ML} = 0.0144$, 
  typically for young ($<$ 6 Myr) clusters, the IGIMF evaluated during 
  the time interval $t,\,t+\delta t$ -- with the (constant) 
  ``star formation epoch" $\delta t = 10$ Myr \citep{WK05a} -- becomes:
  \begin{equation} 
     \label{eq:igimf}
         \xi_{\rm IGIMF}(m,{\rm SFR}(t)) = 
               \int_{M_{\rm ecl,min}}^{M_{\rm ecl,max}({\rm SFR}(t))} 
               \hspace*{-1.8cm} 
    \xi(m\le m_{\rm max}(M_{\rm ecl}))~\xi_{\rm ecl}(M_{\rm ecl})~dM_{\rm ecl}.
  \end{equation}
  A similar timescale $\delta t$ is found by \citet{ESN04} from measuring the 
  offset of {\HII}~regions from the dark spiral arms in a number of galaxies. 
  In our interpretation, $\delta t$ is the time scale within which an ECMF is 
  completely populated. This result is at least qualitatively supported
  by \citet{Elme00a}. By comparing the observational properties of embedded
  clusters and molecular clouds he derives ``that star formation occurs in
  only one or two crossing times''. \citet{HBB01} also argue for short $<$ few
  Myr molecular-cloud life times. And the Antennae \citep{ZF99} have a
  fully developed cluster MF at age of $\sim$ 10 Myr.
  Thus, a lower SFR results in a lower maximum cluster mass which essentially
  reduces the upper mass limit for the stars. Note that
  eq.~\ref{eq:igimf} is the correct ``IMF'' to be used when studying
  the global SFR-properties of galaxies. \citet{Ke83}, for example,
  finds $\alpha = 2.5$ (for m $>$ 1 $M_\odot$) reproduces integrated
  luminosities of star-forming galaxies. Although he refers to this
  ``IMF'' as a ``Salpeter IMF'', it is actually a steeper function
  (with a larger $\alpha$).

\subsection{Model Parameters}
  Because the IGIMF is steeper than Salpeter's IMF, which is often used in 
  chemical evolution studies, its dependence on the SFR can be expected to
  have observable consequences for the chemical enrichment of galaxies. 
  For our evolution models we shall use the following parameters:
  \begin{itemize}
      \item minimum mass of the clusters: 
            $M_{\rm ecl, min} = 5\,\msun$. This corresponds to a
            Taurus-Auriga star-forming ``cluster'' containing a dozen
            stars,
      \item range of ``stellar'' masses: 0.01 to 150 \Msun,
      \item ECMF power-law exponent $\beta = 2$ and also 2.35,
      \item the standard IMF (Eq.~\ref{eq:Kroupa-IMF}).
  \end{itemize}  

\begin{figure}
    \centering
    \includegraphics[angle=-90,width=10cm]{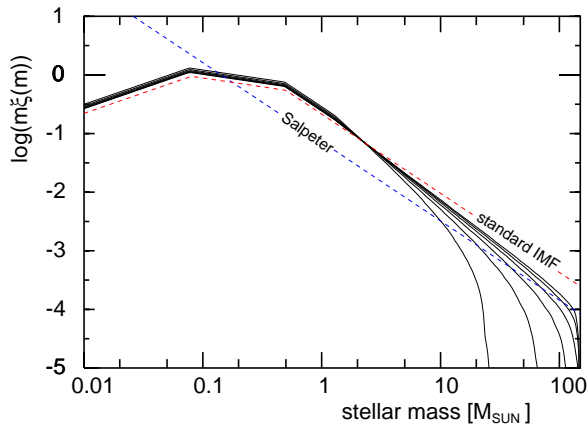} 
    \caption[]{The IGIMFs computed for SFR = 0.01, 
               0.1, 1, 10, and 100 \Msun/yr (solid lines, from bottom
               to top) AND $\beta = 2$, compared to the Salpeter IMF for 
               the entire stellar mass range 0.01 to 150 $M_{\odot}$. All IMFs
               are normalised to have the same number of stars in the
               interval 0.01 to 150 $M_{\odot}$. The standard IMF is
               Eq.~\ref{eq:Kroupa-IMF}.}
    \label{f:imfs}  
  \end{figure}

  The IGIMFs obtained for various SFRs are displayed in Fig.~\ref{f:imfs} 
  in the form of number of stars per unit stellar mass interval (Salpeter's
  IMF has the exponent 2.35). A lower SFR reduces the maximum cluster mass
  and hence the upper mass limit for the stars; this is quite a strong effect.
  One also notes the steepening of the slope, especially at high stellar
  masses. 

\begin{figure}
    \centering
    \includegraphics[angle=-90,width=10cm]{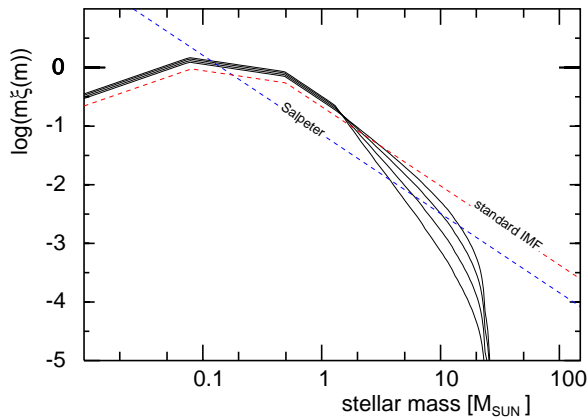} 
    \caption[]{Like Fig. \ref{f:imfs}, but for the same SFR = 0.01 \Msun/yr
               and $\beta = 1.65, 2.0, 2.35$, and 2.7 (solid lines, 
               from right to left).}
    \label{f:imfsb}  
  \end{figure}

  In Fig~\ref{f:imfsb} we show that the slope $\beta$ of the cluster mass 
  function also affects the slope of the IGIMF above stellar masses of
  1 \Msun. For the same SFR, the upper mass limit does not change,
  however increasing $\beta$ further steepens the IGIMF slope. To
  display this dependence, we took a larger range of values for
  $\beta$ than indicated from observations.

  \section{Oxygen yields}
    \label{se:oxy}

   From the IGIMFs we compute the yields of oxygen by integrating over the
   entire stellar mass spectrum the stellar yields weighted with the IGIMFs 
   \citep[cf.~e.g.][]{KA91}. The prescriptions for the stellar nucleosynthesis 
   are: for massive stars we use \citet{TNH96}, including the yields for the
   40 \Msun~star quoted by \citet{TGB98}; beyond 40 \Msun~we assume that the
   mass fractions of freshly produced elements are constant. For low and 
   intermediate mass stars we take \citet{VG97}. This choice is known to give 
   a good match to the chemical evolution of the solar neighbourhood, if one 
   uses a Salpeter IMF \citep[cf.~e.g.][]{TGB98}, because the yield is close
   to the solar oxygen abundance.
   As will be shown by the results below (Fig.~\ref{f:inf1moh}) this choice 
   also gives a good match with the observational data for galaxy masses of 
   $10^{11} \msun$. Uncertainties in the stellar yields would amount to at 
   least a factor of 2 \citep[e.g.][]{HEK00} to be applied in either
   direction.

\begin{figure}
      \centering
      \includegraphics[angle=-90,width=10cm]{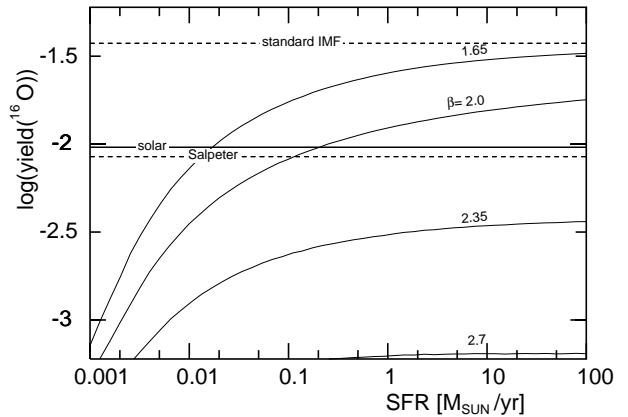} 
      \caption[]{Oxygen yields as a function of the SFR for several values
          of $\beta$. The yields for a Salpeter IMF over all 
          stellar masses and the standard IMF (Eq.~\ref{eq:Kroupa-IMF}) 
          are shown by horizontal dashed lines.}
      \label{f:yo}  
   \end{figure}

   The oxygen yield (Fig. \ref{f:yo}) decreases strongly when the SFR
   is below about 1 \Msun /yr. This mirrors the strong dependence of
   the upper mass limit on the SFR. Test calculations show that the 
   influence of the variation of the IGIMF slope with SFR is of less
   importance. Since for high values of the SFR the IGIMF approaches 
   the standard IMF (cf. Fig. \ref{f:imfs}), the yield converges towards 
   the value obtained with the full standard IMF sampled to the
   maximum cluster mass. The yields also depend on the assumed value
   for $\beta$; larger values result in IGIMFs more strongly cut-off 
   at the high mass end. In the figure we show -- to 
   illustrate this dependence -- a greater range of values for
   $\beta$ than suggested by observations (2.0 ... 2.35; cf. section
   \ref{se:varimf}). 
   A Salpeter IMF for the entire stellar mass range gives an oxygen yield 
   very close to the solar oxygen abundance. As the oxygen yields from massive 
   stars depend on the stellar metallicity only very weakly
   \citep[see][]{WW95}, our results computed for solar composition
   stars also apply to non-solar metallicities.
    
\begin{figure}
      \centering
      \includegraphics[angle=-90,width=10cm]{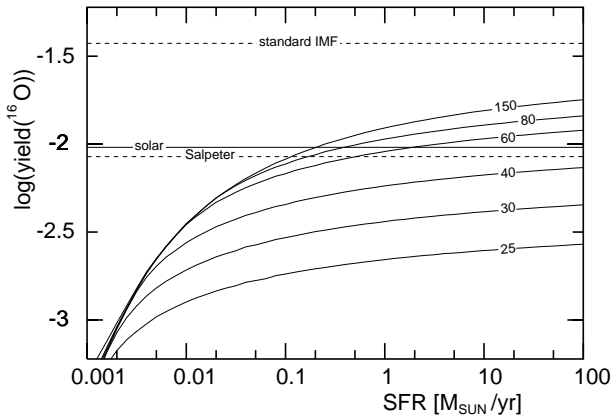} 
      \caption[]{Same as Fig.~\ref{f:yo}, but for $\beta = 2$ and
          several values for the upper mass limit for oxygen production 
          in massive stars (from 150 to 25 \Msun ).}
      \label{f:yo2}  
   \end{figure}

   Stellar oxygen yields of stars as massive as 40 \Msun are less certain 
   due to the poor knowledge about the explosion characteristics \citep{WW95}.
   Furthermore, our adopted nucleosynthesis recipe extrapolates the stellar 
   yields beyond 70 \Msun . We investigated the influence of these
   uncertainties by computing the yields under the assumption that oxygen is
   ejected only by stars below a certain upper mass limit. As depicted in
   Fig.~\ref{f:yo2} the resulting yields are reduced by about a factor of two,
   if oxygen is produced only by stars less than 40 \Msun . The dependence on
   the SFR becomes somewhat less steep for low values of the SFR.
  
   If all galaxies had the same age of about 10 Gyr, galaxies with masses 
   between $10^7$ to $10^{11} \msun$ have time-averaged SFRs ranging from
   0.001 to 10 \Msun/yr (cf.~Fig.~\ref{f:yo}). Since over this interval, the 
   oxygen yield increases by about a factor of thirty, we may thus 
   expect that the SFR-dependent IGIMF causes observable effects in the 
   chemical evolution of these systems.

\section{Chemical Evolution Models for Galaxies of Different Masses}
  \label{se:chevo}

   In order to study the effects of the SFR-dependent IGIMF on the present-day
   gas abundances of galaxies of different masses, we compute a series of
   illustrative chemical evolution models to display the principal effects 
   and show their magnitude. These models use simple parameterisations, and
   thus are not designed to match the exact time histories of the galaxies,
   however we concentrate on those properties -- metallicity and gas fraction
   -- in which chemical evolution models are known to be less sensitive to
   modelling details. Thus, the results retain a certain degree of general
   validity.

  \subsection{Physical Assumptions}

   Each galaxy is treated as a one-zone model, dealing with the gas return 
   and metal production of stars of different masses in detail and taking into
   account the finite lifetimes of the stars. The ejected gas and metals are 
   assumed to be mixed perfectly and instantaneously with the gas in the
   model, as is done in the usual chemical evolution studies
   \citep[cf.~e.g.][]{TA71,Ma86}. As described before, the stellar
   nucleosynthesis prescriptions are from \citet{TNH96} for massive
   stars (type II supernovae) and \citet{VG97} for low and
   intermediate-mass stars. 

   At each time-step the IGIMF is selected in accordance with the
   current value of the SFR, using a pre-computed lookup table. The
   SFR has the same form as used by e.g. \citet{MF92},
   \begin{equation}
        \label{e:sfr}
        {\rm SFR}(t) = {M_{\rm gas}(t)+M_{\rm stars}(t)\over \tau_{\rm SFR}} 
               \left({M_{\rm gas}(t) \over M_{\rm gas}(t)+M_{\rm
               stars}(t)}\right)^x,
   \end{equation}
   which depends on the current masses of gas, $M_{\rm gas}(t)$, and stars, 
   $M_{\rm stars}(t)$, and thus on the current values of total mass and gas 
   fraction. The parameter $x$ is taken as unity, although test
   calculations are done with other values, giving very similar results,
   provided the present-day gas fraction remains the same, as one should
   expect for closed-box and infall models.  
   The star formation time scale $\tau_{\rm SFR}$ is assumed to be the same for
   all galaxy masses; several values shall be taken, which will thus determine
   the present gas fraction. In later models, we shall allow
   $\tau_{\rm SFR}$ to vary with galaxy mass.

   The galaxy is either treated as a closed box with all its initial mass in 
   the form of metal-free gas, or as an infall model where the primordial gas 
   is accreted with a rate that decreases exponentially in time:
   \begin{equation}
        \label{e:infall}
        \dot{M}(t) =  M_{\rm gal}   {\exp(-t/\tau_f) \over 1-\exp(-T/\tau_f) }, 
   \end{equation}
   with the time scale $\tau_f = 5$ Gyr, similar to \citet{Ch80},
   \citet{MF92} and \citet{TGB98} for models of the solar
   neighbourhood. We assume all the mass of the gas to be accreted
   within the age $T= 13$ Gyr of the galaxy. It is worth pointing out
   that with this kind of monotonic gas infall, the oxygen abundance
   remains close to the values from a closed-box model of the same gas
   fractions \citep{KE99}. Thus, the choice of the infall time scale
   is of minor importance in the context of the primary element oxygen
   ejected with very little time delay. 

   For the numerical calculations we use a grid of 190 logarithmically
   spaced stellar masses and 500 time steps to cover 13 Gyr.

  \subsection{Closed-Box Models} 
   
   We first compute a series of closed-box models.  The star formation 
   time scale is taken to be a fraction or multiple of the age of the galaxy: 
   $\tau_{\rm SFR} = T/a$. The factor $a$ is assumed to take values of
   0.5, 1, 2, 3, and 4, i.e. time scales of 26, 13, 7, 3.5 and 1.75 Gyr. 
   These five series of models with different galactic masses span a rather 
   wide range of present-day gas fractions,
   \begin{equation}
        f_{\rm gas} (T) =  {M_{\rm gas}(T) \over M_{\rm gas}(T)+M_{\rm
        stars}(T)}. 
   \end{equation}

\begin{figure}
      \centering
      \includegraphics[angle=-90,width=10cm]{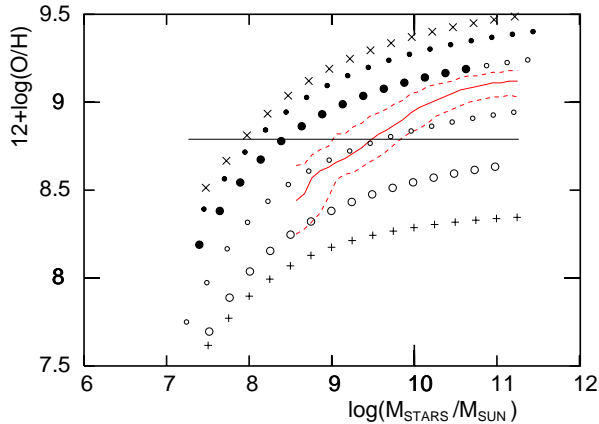} 
      \caption[]{The relation between oxygen abundance and
                 stellar mass of galaxies. Dashed lines indicate the
                 16, 50, and 84 percentiles of the observed metallicity 
                 distribution by \citet{THK04}.
                 The symbols are results from closed-box models:
                 From top to bottom, the model series (with $\beta = 2$)
                 correspond to the factor $a$ in the SFR 
                 to 4, 3, 2, 1, and 0.5.
                 The symbols indicate the present-day gas
                 fraction: large circles $>0.5$, 
                 small circles $0.5 ... 0.25$,
                 large filled circles $0.25 ... 0.15$,
                 small filled circles $0.15 ... 0.1$,
                 and crosses $<0.1$. 
                 The plus-signs show the model series
                 with $a=1$ but $\beta=2.35$. 
                 The full horizontal line refer
                 to a model grid obtained with a constant Salpeter
                 IMF between 0.01 and 150 \Msun . }
      \label{f:simmoh}  
   \end{figure}

   In Fig. \ref{f:simmoh} we compare these models with the data of
   \citet{THK04} for the relation between stellar mass of the galaxies and 
   gas-phase oxygen abundance. The computed relations are somewhat 
   flatter than the observed relation, but models with present-day gas 
   fractions between 25 and 50\% are indeed well suitable to explain the 
   observed metallicities, which correspond to $a \approx 1$.
   For comparison we show one model series done with a constant 
   Salpeter IMF. Since in these models the oxygen abundance is a function of
   only the gas fraction which is determined by the star formation time scale,
   each of the five model-series give an oxygen abundance independent of the 
   galactic mass. The model shown is done with $a=1$, the other models
   result in horizontal lines at different oxygen abundances, ranging
   from 8.5 for $a=0.25$ to 9.4 for $a=4$.  
  
\begin{figure}
      \centering
      \includegraphics[angle=-90,width=10cm]{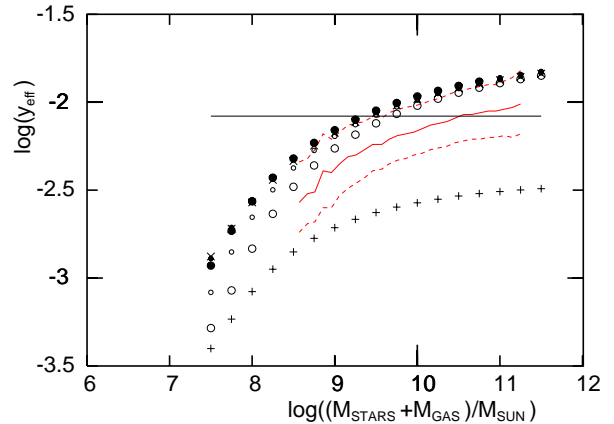}
      \caption[]{The relation between the effective O yields and the total
                 galactic mass: dashed lines indicate the
                 16, 50, and 84 percentiles of the observed
                 distribution by \citet{THK04}. The symbols
                 denote the same sequences of closed box models of 
                 Fig. \ref{f:simmoh}. The horizontal line refers to
                 models with a constant Salpeter IMF. The plus-signs
                 show the model series with $a=1$ but $\beta=2.35$.}
      \label{f:simmyeff}  
   \end{figure}

   \citet{THK04} also determine the relation between
   total galaxy mass $M_{\rm gal} = M_{\rm gas}(T)+M_{\rm stars}(T)$
   and the effective oxygen yield, which they compute from current
   oxygen abundances and gas fractions by assuming a closed box model,
   \begin{equation}
       y_{\rm eff} = -{Z(^{16}{\rm O}) \over \ln f_{\rm gas}(T)}.
   \end{equation} 
   Our closed box model sequences (Fig.~\ref{f:simmyeff}) reproduce 
   the slope of the relation remarkably well, although being higher 
   than the median observed relation by 0.1 dex. This is in strong 
   contrast to models with a constant IMF which all have the same 
   effective yield for all galactic masses and all star formation 
   time scales.

  \subsection{Infall Models}

\begin{figure}
      \centering
      \includegraphics[angle=-90,width=10cm]{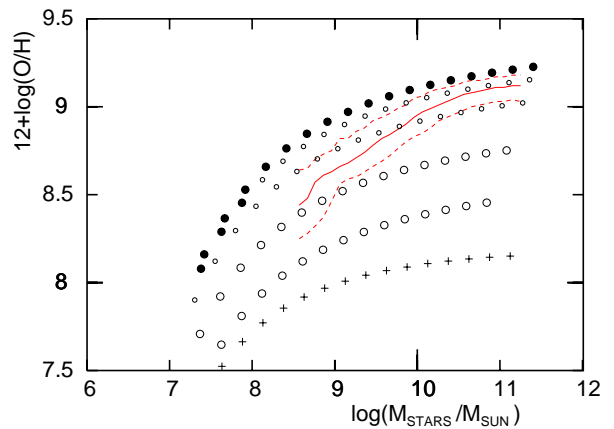}
      \caption[]{Same as Fig. \ref{f:simmoh} but for models
                 with an exponentially decreasing infall
                 of primordial gas.}
      \label{f:infmoh}  
   \end{figure}

   Motivated from chemical evolution models for the solar neighbourhood,
   we consider continuous infall of primordial gas
   \citep[e.g.][]{MF92, TGB98, NO06}. The infall rate is assumed to decrease
   exponentially  in time with a time constant of 5 Gyr. As can be
   seen from Fig.~\ref{f:infmoh}, one obtains larger gas fractions
   and lower metallicities. Thus, models with $a \approx 2$ are quite
   suitable. We note that infall does not change the slope of the
   relation which remains slightly too flat. 

   As expected for models with monotonically decreasing infall, the effective 
   yields are only slightly (less than 0.1 dex) lower than found from the 
   closed-box models. Hence, they also match remarkably well the shape of 
   the observed median relation.

\begin{figure}
      \centering
      \includegraphics[angle=-90,width=10cm]{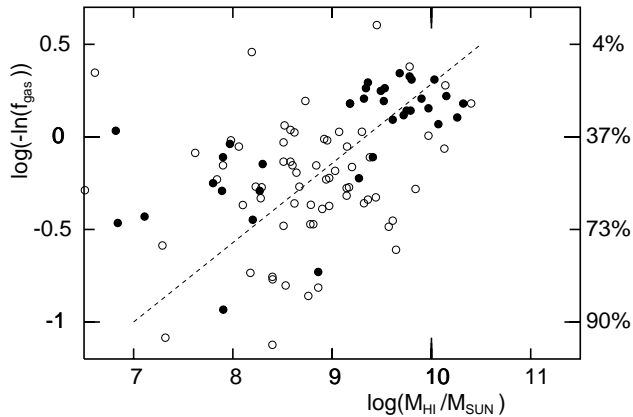}
      \caption[]{The observed relation of hydrogen mass
               and gas fraction \citep[adapted from][where
               details may be found]{KH05}. The filled dots mark objects
               with gas fractions determined by \citet{Ga02} that
               account for both atomic and molecular content. 
               The dotted line is an assumed relation, used for
               Fig.~\ref{f:deltafz}.}               
      \label{f:mhifgaso}  
   \end{figure}

\begin{figure}
      \centering
      \includegraphics[angle=-90,width=10cm]{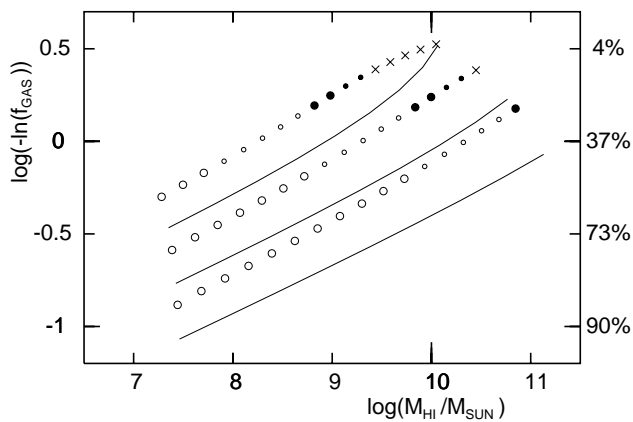}
      \caption[]{The relation between present gas mass and gas
                 fraction from models with gas infall of 5 Gyr 
                 time scale. The central series of symbols refers to
                 star-formation time scales which depend on the galaxy's 
                 mass following Eqn. \ref{e:timesfr}. The SFR 
                 time scales are halved in the upper series and doubled
                 for the lower sequence. The meaning of the symbols is
                 as in Fig. \ref{f:simmoh}. The three full lines are for
                 similar models, but computed with a constant Salpeter 
                 IMF.}
      \label{f:inf1mhifgas}  
   \end{figure}

   Since a general trend is observed that massive galaxies have smaller gas
   fractions than dwarfs \citep[e.g.][]{KH05}, we construct another
   set of models in which the star formation time scale is assumed to
   decrease with increasing galactic mass $M_{\rm gal}$ in such a way, 
   \begin{equation}
       \label{e:timesfr}
       \tau_{\rm SFR} = 20\ {\rm Gyr} 
               \ \ \cdot \left({M_{\rm gal}\over 10^8 \msun}\right)^{-0.3} 
   \end{equation}
   that the observed data (Fig. \ref{f:mhifgaso}) are reasonably well
   matched. 
   In Fig.~\ref{f:inf1mhifgas} we show three model sequences,
   done with the above time scale, as well as with half and twice its value.
   For comparison, we also compute three model sequences with a constant
   Salpeter IMF. A similar dependence of the SFR time scale on galactic mass
   ($ \tau_{\rm SFR} = 50\,{\rm Gyr} \cdot  (M_{\rm gal}/ 10^8 \msun
   )^{-0.25}$) gives a satisfactory match to the observations.

\begin{figure}
      \centering
      \includegraphics[angle=-90,width=10cm]{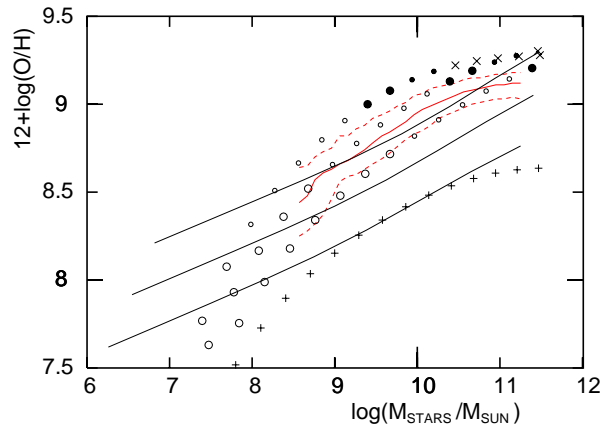}
      \caption[]{Similar to Fig.~\ref{f:simmoh} but for
                 models with exponentially decreasing infall
                 of primordial gas and with a star formation
                 rate depending on the galactic mass (Eqns.~\ref{e:sfr} 
                 and \ref{e:timesfr}). The three series of models 
                 are as in Fig.~\ref{f:inf1mhifgas}. The full
                 curves refer to a model computed with
                 a constant Salpeter IMF, while the plus-signs show
                 the model series with $a=1$ but $\beta$ = 2.35.}
      \label{f:inf1moh}  
   \end{figure}

   The resulting model sequences match very well the distribution 
   observed by \citet{THK04}, as is seen in Fig. \ref{f:inf1moh}. 
   It is worth noting that the agreement is excellent both in terms 
   of slope and metallicity. 

\begin{figure}
      \centering
      \includegraphics[angle=-90,width=10cm]{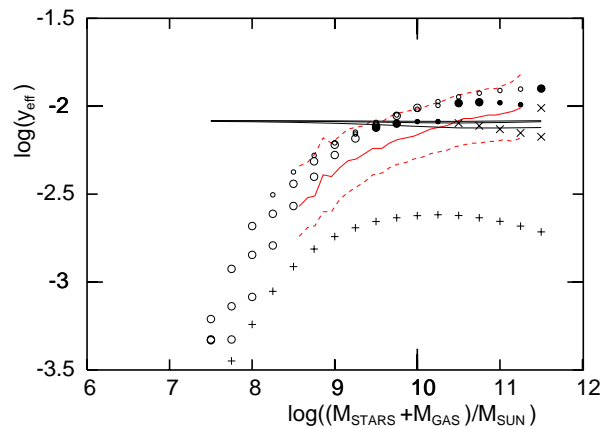}
      \caption[]{Similar to Fig. \ref{f:simmyeff}, but for the infall
                 models of Fig. \ref{f:inf1mhifgas}. The symbols
                 are the same as in Fig. \ref{f:simmoh}. The  
                 nearly horizontal curves are from the models
                 computed with a constant Salpeter IMF.}
      \label{f:inf1myeff}  
   \end{figure}

   The effective yields (Fig.~\ref{f:inf1myeff}) also match the
   observed relation very reasonably, although the slope among the massive 
   galaxies is flatter and even turns down, and near $10^{9.5} \msun$ the 
   yields are about 0.1~dex too large.

   The adopted infall time scale is not a critical parameter for this
   match with observations: monotonically decreasing infall leads to
   but minor deviations from the closed-box model properties, in 
   particular the relation between gas fraction and metallicity.
   As is evident from Figs.~\ref{f:simmoh} and \ref{f:infmoh}, the
   metallicity in more massive galaxies are decreased by about 0.2~dex;
   but it is worth emphasising that the locus of the galaxies with
   the {\it same gas fraction} has hardly changed at all. The same
   holds true if one chooses the infall time scale to be anywhere 
   between about 1 and 10 Gyr. 

   The models with a constant IMF do not provide such a good match: While
   the assumed relation between SFR time scale and galactic mass could account 
   for an increase of metallicity with galactic mass, our simple recipe 
   cannot reproduce the shape of the observed relation (Fig.~\ref{f:inf1moh}). 
   Furthermore, these models fail completely to match the
   mass-effective--yield relation, as their yields are nearly
   independent of galactic mass (Fig.~\ref{f:inf1myeff}). Thus the
   dependence of effective yield on galaxy mass might be more crucial
   in tracing the origins of the mass--metallicity relation. 

  \subsection{Models with starbursts}
    \label{se:bursts}

\begin{figure}
      \centering
      \includegraphics[angle=-90,width=10cm]{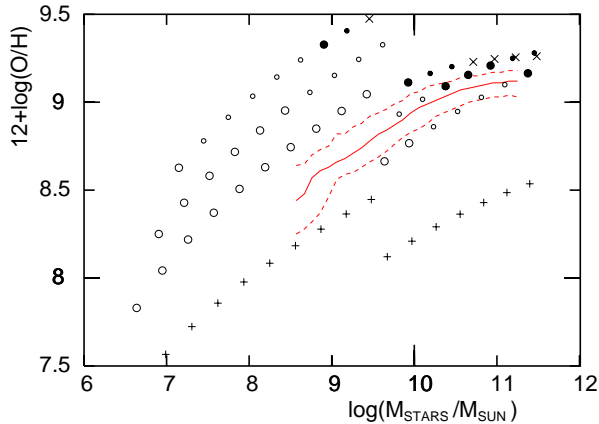}
      \caption[]{Like Fig.~\ref{f:inf1moh} but assuming that for galaxies with
                 a {\it total} mass below $10^{10} \msun$ 
                 star formation occurs in bursts of 
                 100 Myr duration every 1 Gyr. During a burst the
                 SFR is 30 times higher than the average rate.}
      \label{f:binf1moh}  
   \end{figure}

\begin{figure}
      \centering
      \includegraphics[angle=-90,width=10cm]{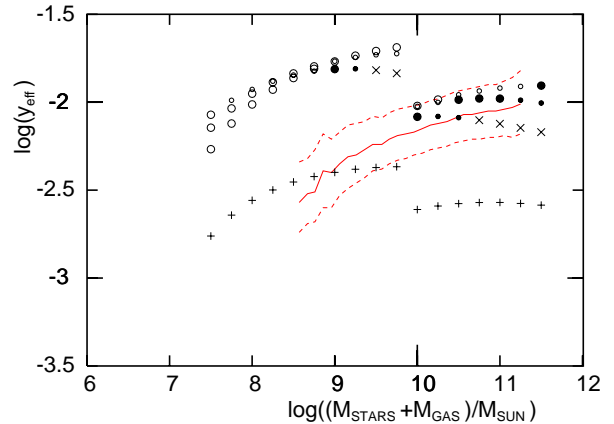}
      \caption[]{Like Fig.~\ref{f:inf1myeff} but assuming that
                 for galaxies with a {\it total} mass below $10^{10} \msun$
                 star formation occurs in bursts of 
                 100 Myr duration every 1 Gyr. During a burst the
                 SFR is 30 times higher than the average rate.}
      \label{f:binf1myeff}  
   \end{figure}

   Among dwarf irregular galaxies the star formation history is known to
   show strong temporal variations \citep[e.g.][]{Gr01}. How does the
   presence of star bursts influence the results? Let us consider the
   extreme case, that all star formation occurs in short periods with
   long intervals of no star formation. We shall assume that in
   galaxies with less than $10^{10} \msun$ stars are formed in a
   burst lasting 100 Myrs every 1 Gyr. In order to obtain the same
   present gas fraction as in a model with continuous star formation,
   the SFR during the short bursts must be 30 times higher than the
   average rate. The resulting IGIMF contains a larger fraction of 
   massive stars, and thus the average metal yield will be higher than 
   under continuous star formation. Thus, strong starbursts lead to both 
   higher oxygen abundances
   (Fig.~\ref{f:binf1moh}) and higher effective yields
   (Fig.~\ref{f:binf1myeff}), as shown in this somewhat extreme model. 
   One notes that since for these models the SFR is 
   higher, the variations of the upper stellar mass limit and the
   IGIMF slope should be milder. As Fig.~\ref{f:binf1myeff} shows,
   the variation of the effective oxygen yield is less than in the
   models with continuous star formation (cf. Fig.~\ref{f:inf1myeff}).
   In the oxygen abundances, this change is much less apparent,
   because the assumed dependence of the SFR time scale on
   galactic mass (Eqn.~\ref{e:timesfr}) contributes fairly strongly to 
   the mass--metallicity relation (cf. Fig.~\ref{f:inf1moh}).
   
   We emphasise that this schematic modelling of star formation bursts 
   serves to show the influence on the results of a rather extreme 
   burst-type history of star formation, irrespective of the physical 
   processes which could form the base for such a behaviour. The 
   parameters were chosen to provide a clear demonstration of the
   effects. Nonetheless, the results predict that with a SFR dependent
   IGIMF one would expect that galaxies which had a strongly fluctuating
   star formation history would tend to have higher metallicities
   and higher oxygen yields.
 
  \section{Discussion}
   \label{se:discuss}

   \subsection{Chemical evolution models}

   The consequences of the IGIMF depending on the SFR is primarily that
   the oxygen yield is higher in environments -- or times -- where the
   SFR is high. Thus, more massive galaxies are expected to have higher
   yields and thus have a tendency to have higher metallicities.
   This basic implication is quantified in the chemical evolution models
   presented in Sect.~\ref{se:chevo}. 
   \begin{itemize}
      \item The closed-box models exhibit mass-metallicity and 
            mass--yield relations which are very similar to what is 
            observed, without the need to invoke additional physics.
      \item Adding infall of primordial gas modifies both relations only
            slightly.  The presence of gas infall or the true history
            of the infall -- as long as it decreases monotonically in 
            time -- are not crucial for this comparison.
      \item If the observed trend between galaxy mass and present gas fraction
            is matched by adopting a dependence of the star 
            formation time scale on galaxy mass, both observed relations of
            \citet{THK04} can be reproduced very well. 
            This supplementary assumption leads to a further
            improvement of the fit of the mass--metallicity relation,
            but it is not an essential ingredient, because it alone
            cannot account for the mass--effective yield relation.
   \end{itemize}
   Thus, the main contribution to the agreement with the \citet{THK04} 
   data is the variable IGIMF, and not the details of the model.

   \subsection{Star formation history}

   We have chosen a rather simple recipe for the SFR and adopted a dependence
   of its time scale on galaxy mass on the basis of matching the observed trend
   in the current gas fractions on galaxy mass. Chemical evolution models with 
   constant yield but gas infall remain rather close to the relation of the 
   closed-box model, unless subjected to massive and rapid inflow of
   gas \citep{KE99,KH05}. Because oxygen is an element whose enrichment 
   occurs with negligible time delays, its present abundance is predominantly 
   determined by the gas fraction, but irrespective of  the time dependence of 
   the SFR.

   This applies also to our models with a variable IMF and {\it continuous}
   star formation: in massive galaxies with their high SFR the yield is rather
   insensitive to changes in the SFR (Fig.~\ref{f:yo}). The instantaneous
   yield changes very little in the course of time, as evident e.g. in
   Fig.~\ref{f:simmyeff}: only for low galaxy masses do the models
   with different $a$ give different effective yields, because during
   its time evolution the low SFR results in a strong change of the
   yield. Although in the models the IGIMF depends both on time and
   galaxy mass, it is the dependence on the latter parameter which
   clearly dominates. Therefore one has a direct mapping of the
   SFR--yield relation into the mass--effective-yield relation, as the
   close similarity of the curves in Figs.~\ref{f:yo} and
   \ref{f:simmyeff} shows. Furthermore, models with a constant star
   formation time scale have the same gas fraction for all galaxy
   masses, hence the closed-box relation $Z = - y \ln(f_{\rm gas})$
   provides a common scaling factor between yield and abundance
   (Fig.~\ref{f:simmoh}).

   Therefore, as far as a continuous star formation is concerned, a comparison
   of its history predicted by our models with observations does not
   constitute a  critical test of their validity or of the variable IMF
   concept. The presence of burst-like forms of star formation and the
   consequences in conjunction with a variable IMF are addressed later in this
   section. But it worth noting that the time scale for star formation in the
   universe is found to be about 7~Gyr \citep{GBB03, BCW04} and that models
   with such a value (i.e. $a=2$) match quite well the observed
   mass--metallicity relation (Figs.~\ref{f:simmoh} and  \ref{f:infmoh}). The
   reason is the already mentioned simple mapping of the SFR--yield relation
   into the mass--metallicity relation: With a value of 7~Gyr a closed-box
   model with linear SFR ($x=1$) gives a present gas fraction of $f_{\rm
   gas}^0= 0.13$. Since the gas fraction and thus the SFR decrease
   exponentially, a galaxy of $M_{\rm gal} = 10^{10} \msun$ has an average
   SFR of  $M_{\rm gal} (1- f_{\rm gas}^0)/13 {\rm Gyr}$ $= 0.67$ \Msun
   /yr. From Fig.~\ref{f:yo} one obtains a yield slightly above solar
   value. The closed-box relation then gives the oxygen abundance $Z = - y
   \ln(f_{\rm gas})$  as about twice the yield, as is seen in
   Fig.~\ref{f:simmoh}.

   Observational data are available to compare with our adopted relation
   between star formation time scale and galaxy mass: \citet{HPJ04} determine
   the stellar populations in nearly 100000 galaxies from the continua
   observed in the Sloan Digital Sky Survey and deduce the time dependence of
   the SFR in galaxies as a function of their mass. While galaxies with masses
   of less than $10^{10} \msun$ have star formation increasing slightly with
   time, which might indicate of a very slow infall of gas, the SFR for
   e.g. $10^{11} \msun$ initially is about 100 times higher and declines
   rapidly, with a time scale of about 1 ... 2~Gyr. This behaviour also
   implies that today's low mass galaxies remain richer in gas than the
   massive systems. 
   Our recipe gives a similar value of 2.5 Gyr for $10^{11} \msun$, but 
   with 5 Gyr for $10^{10} \msun$ the gas consumption is too rapid. For lower
   masses, our time scale increases and thus the star formation history is 
   dominated by the infall which results in a nearly constant or increasing
   rate. This can be considered consistent with the data of \citet{HPJ04}
   which lump all masses below $10^{10} \msun$ in a single bin.
       
   In a sample of the same size \citet{BCW04} find that for masses below about
   $10^{10} \msun$ the SFR has a tight relation with the stellar mass in a
   galaxy (their Fig.~17). However, the high ratios of present and average past
   SFR confirm that star formation is dominated by bursts, and therefore its
   current value is not necessarily representative for the average rate of gas
   consumption. Therefore, for more detailed modelling of these
   galaxies the SFR recipe would need to include a proper description
   of the burst-type nature of star formation.

   We have refrained from doing this in the present work, but in Section
   \ref{se:bursts} we illustrate the effects of starbursts on the results. In
   this extreme model with ${\rm SFR}/<{\rm SFR}>$ = 30, which is
   about ten times larger than the largest value shown in Fig.~24 of
   \citet{BCW04}, the effective yield and the oxygen abundance for
   galaxies with $< 10^{10} \msun$ are enhanced by about
   0.4~dex. Since these galaxies have ${\rm SFR} < 0.7 \msun$/yr, we
   see from Fig.~\ref{f:yo} that the yield increases by 0.3
   ... 0.4~dex for an increase of a factor of 30 in the SFR, and  0.1
   ... 0.2~dex for a factor of 3. Thus the enhancements would be less
   than half of those shown in Figs.~\ref{f:binf1moh} and
   \ref{f:binf1myeff}. The results of \citet{BCW04} would imply that
   while at masses near $10^{10} \msun$ there would a negligible
   change from our results but that due to the effects of starbursts
   the effective yields and the oxygen abundances would progressively
   be enhanced for lower masses, amounting to about 0.2~dex at $10^{8}
   \msun$. The low-mass branch in both relations would be somewhat
   flatter than predicted from our continuous star formation
   models. However, this difference remains comparable with the
   uncertainties of the stellar nucleosynthesis prescriptions and of
   the observational data, to be discussed in the subsequent sections.

   \subsection{Oxygen abundances}

   The absolute values of abundances and yields from the models depend 
   on the adopted recipes for the stellar nucleosynthesis, which may differ
   in the underlying stellar physics, such as the treatment of convection,
   the nuclear cross sections, and the explosion characteristics. For example,
   the stellar oxygen yields from \citet{TNH96} and \citet{WW95} can differ 
   for the same stellar mass by as much as a factor of two, as shown 
   by \citet{TGB98}. Other effects, such as stellar mass loss \citep{M92} 
   and rotation \citep[e.g.][]{MM02} also have a strong influence on the
   stellar yields. In view of these uncertainties as well as of those in the
   observational data, which are addressed below, the agreement of the
   absolute abundance values between our models and the data of \citet{THK04}
   should be regarded but as a welcome aspect. The essential feature of the
   variable IMF models is the agreement in the overall shape of both
   relations, already apparent in the closed box models.

   \subsection{Galaxy formation}

   One basic assumption in the models is that the star formation history
   of a galaxy is governed by an overall time scale. If galaxies form as
   isolated entities, one could expect such a behaviour. If -- to take the
   hierarchical galaxy formation to a limiting case -- all stars were formed
   in smaller substructures and in the same way, from which galaxies are 
   built by mere additions, one should expect that all galaxies should 
   have the same present gas fraction. The variation of gas fractions 
   and present star formation among spiral and irregular galaxies gives 
   evidence that a substantial star formation must have taken place during 
   or after the assembly of a galaxy. In the first place, a more massive
   galaxy would have undergone a larger number of star formation episodes
   which could well result in an overall relation between star formation
   and present galactic mass. What fractions of stars are formed in either of
   these two modes is not yet clear. \citet{CJC06} finds that most of 
   today's massive galaxies probably have accreted the majority of their mass
   in only 4 or 5 major mergers in the early universe. As the merging objects
   already must have been quite massive ($\sim 2 \times 10^{10}~M_{\odot}$ in
   baryons) and young, the resulting IGIMF would be quite similar to the final
   IGIMF as the change in the IGIMF is most prominent for galaxies below
   $10^{10}~M_{\odot}$ \citep{WK05a}. \citet{UWG96} and \citet{VIS04}
   have shown that the Milky Way stellar population has abundance
   patterns that cannot be derived from accreted satellites, unless
   they had significantly different stellar populations than today's
   dwarf spheroidal galaxies. This, however, would be unphysical,
   because $\le 10^7 M_{\odot}$ satellites would not be able to
   chemically enrich to the Milky Way values.

   If a major portion of the present stellar population was formed in
   smaller substructures, probably with a high SFR, or if each merging of
   a substructure results in an episode of enhanced star formation,
   one should expect higher values for the average effective yield and
   the metallicity -- as demonstrated in the models with bursty star 
   formation histories. Thus the variation of the metallicities could be
   less than computed from our models.

   Our models -- closed box and infall -- thus constitute an upper
   limit of the effects due to a variation of the IGIMF on the SFR.
   
   \subsection{Galactic Outflows}

   The effective yields can further be reduced by the loss of metals due 
   to galactic winds. While the details of the effects of supernova explosions
   on the chemical evolution are quite complex \citep[e.g.][]{MF99,RHAM06}, one
   may distinguish two simplified cases:
   
   Some part of the metal-rich ejecta freshly produced by the massive 
   stars are directly expelled from the galaxy. How large this fraction 
   is depends on the details of the explosion, the multi-phase structure
   of the ISM, and the location in the galaxy \citep[see e.g.][]{FHY03}, which
   makes it difficult to give reliable estimates. It 
   is obvious, however, that the effective yield could be reduced to
   arbitrarily small values, as it might be possible that {\it all} stellar
   ejecta leave the galaxy. Indeed, this issue is unclear, as
   \citet{STT98,STT01} show that even massive galaxies would loose all
   their metals through outflows if the \citet{MF99} estimates are adopted.

\begin{figure}
      \centering
      \includegraphics[angle=-90,width=10cm]{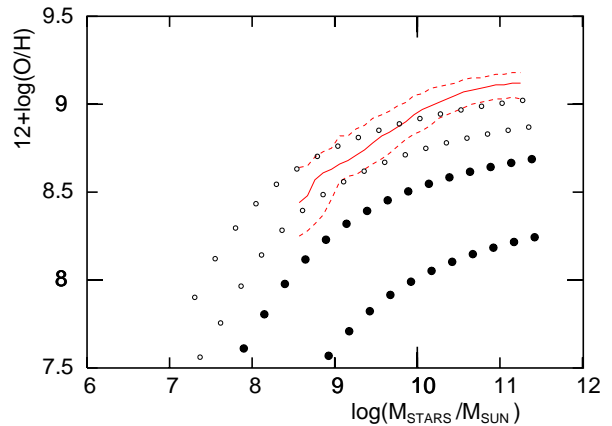}
      \caption[]{Same as Fig. \ref{f:infmoh}, but with additional gas outflows.
                 The star formation time scale is 7.5 Gyr ($a=2$), independent 
                 of galaxy mass. The top series of symbols (open ones) are
                 models without outflows; for the ones below the ratio of
                 outflow rate to SFR is taken to be constant at 1, 2, and
                 5. The symbols indicate the gas fractions, as in
                 Fig. \ref{f:simmoh}.}
      \label{f:windmoh}  
   \end{figure}

   The other case is that the galaxy loses part of its ISM, following
   each event of star formation. Thus, the outflow rate would be expected to
   be proportional to the SFR. In Fig. \ref{f:windmoh} we illustrate how
   the infall models of Fig. \ref{f:infmoh} are altered by outflows of this
   type: if the ratio of outflow and star formation rates is 1, 2, or 5,
   the metallicities are reduced by 0.15, 0.3, and 0.75~dex, respectively.
   The effective yields are reduced by 0.3, 0.5, and 1.1~dex. Since in these
   models we use a constant star formation time scale, the gas fractions do
   not vary with galaxy mass. Thus, both the mass--metallicity and mass--yield
   relations are essentially modified by a multiplicative factor. Note that
   the outflows also will lower the gas fractions. 
   
\begin{figure}
      \centering
      \includegraphics[angle=-90,width=10cm]{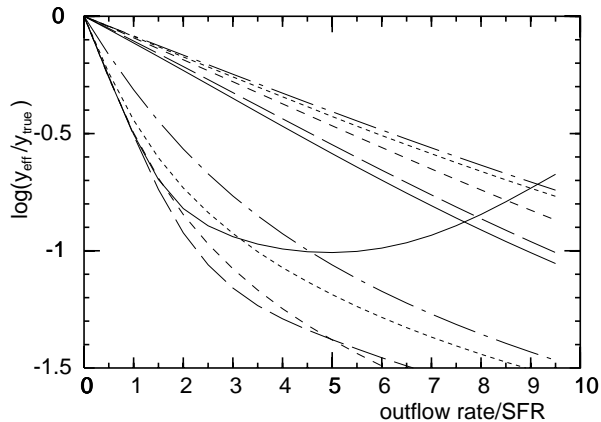}
      \caption[]{The reduction factors of the effective yield below the
                 true yield for models where the outflow rate is a constant
                 multiple of the SFR. The upper set of curves refer to models
                 with low SFR ($a=0.5$), the lower one to those with $a=4$.
                 The full curves are without infall, the dashed curves are
                 with infall time scales of 1, 5, and 10~Gyr (shortest
                 dashes). The dash-dotted curves are for an infall time scale
                 of 5 Gyr, but with a quadratic SFR ($x=1$).}
      \label{f:winyield}  
   \end{figure}

   The reduction factor of the effective yield depends strongly on the star
   formation time scale and the ratio of outflow rate and SFR, but much less
   on the infall time scale, and the exponent $x$ of the SFR, as presented in
   Fig. \ref{f:winyield}. Since suitable models have star formation time
   scales of about 7 Gyr ($a=2$), we may take from these schematic models 
   that in order to cover the difference of about 1~dex in oxygen abundances 
   and in the yields between massive and dwarf galaxies, the ratio of outflow 
   rate and SFR needs to vary only between 0 and about 5. This number is 
   likely to be near the lower end of this range, because stellar ejecta may
   also escape directly from dwarf galaxies.

   The combination of gas outflows and a variable IGIMF yields a
   mass--metallicity relation that is steeper than the one produced by either
   process alone. On the other hand, the presence of strongly fluctuating star
   formation in dwarf galaxies leads to higher yields -- as shown in Section
   \ref{se:bursts} -- which counteracts the effects of the variable IGIMF and
   thus gives a flatter mass--metallicity relation. Thus, chemical evolution
   models with parametrised recipes would not be sufficient to discriminate
   between the three aspects. To further quantify this is beyond the scope of
   the present work, which focuses on the consequences of a variable IGIMF on
   the chemical properties of galaxies.

   \subsection{Observational data}

   To compare our models with the observed mass--metallicity relation
   we use the recent observational data of \citet{THK04} which constitutes
   a rather large and also homogeneous set of data.
   While our study is not the place to discuss in depth all uncertainties and
   limitations of this data set, several points need to be addressed: 
   the abundances are estimated from the strong lines, thus without measuring
   the electron temperature in the ionised gas of each object which would
   require the observation of the faint diagnostic lines. While this can be
   used with calibration with photoionisation models to obtain oxygen
   abundances, direct comparison with determinations with measured electron
   temperatures have shown that strong line methods may overestimate the
   oxygen abundance by as much as 0.5 dex \citep[cf.][]{KBG03}. Furthermore,
   the data pertain to the central regions of the galaxies which are likely to
   be of higher metallicity than the averaged value which the models are
   designed to compute. 
   The computation of the effective yields requires the knowledge of the gas
   content of each object, which would need a determination of the mass in
   stars and gas in neutral, ionised, and molecular form. The derivation of
   the gas fraction is further made more complicated in that such a measure
   should exclude any gas that does not take part in the star-gas cycle of
   evolution. Such details of the galaxies are not available for the objects
   studied by \citet{THK04}; they can determine the gas fraction but by the 
   indirect means of determining the gas surface density from the H$\alpha$
   luminosity via the relation between SFR and gas surface density by
   \citet{Ke98}. But this verifies that the obtained relation between
   galaxy mass and effective yields is in good agreement with the data
   from a number of individual galaxies available in the literature.

\begin{figure}
      \centering
      \includegraphics[angle=-90,width=10cm]{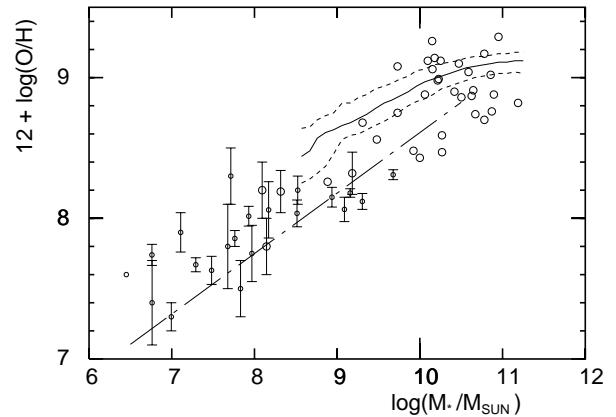}
      \caption[]{Comparison of the stellar-mass--metallicity relation 
                 determined by \citet{THK04}
                 with abundances for individual gas-rich galaxies
                 from the sample of \citet{KH05}. Large circles indicate
                 spiral galaxies, small filled circles are irregulars.
                 The dash-dotted line is a fit to Garnett's (2002)
                 data on oxygen abundances and \HI~mass.}
      \label{f:mstaroho}  
   \end{figure}

   We have compared the relations of \citet{THK04} to data from individual
   gas-rich galaxies, taken from the literature \citep[for details
   see][]{KH05} for different objects than used in their comparison.
   In most of these galaxies, abundances are determined by measuring the
   electron temperature of the \HII~gas. As shown in Fig.\ref{f:mstaroho}, 
   these individual data show oxygen abundances systematically lower
   by about 0.3~dex, which is in line with the findings of
   \citet{KBG03} in M~101.  This is also in agreement with the results
   by \citet{YLH06} who compared the oxygen abundances for 531
   spectra of star-forming galaxies from the SDSS, based  on the
   direct determination of electron temperatures from the [O III]
   5007/4363 line ratio with the Bayesian estimates via strong lines
   by \citet{THK04}. They  find that in about half of the
   sample the strong-line technique overestimated the abundance by 0.3
   dex (their Fig. 2), which is correlated to the N/O abundance
   ratio. In the framework of our models, such an 
   offset could be accounted for by a slightly higher value of $\beta
   = 2.2$, for instance. Such a change would still be well within the
   range of acceptable values for this parameter. Merely for
   comparison, we also plot the relation of \HI~mass and metallicity
   by \citet{Ga02}.

\begin{figure}
      \centering
      \includegraphics[angle=-90,width=10cm]{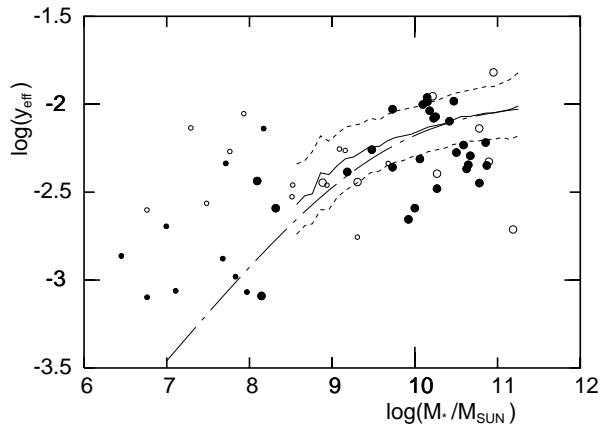}
      \caption[]{Comparison of the stellar-mass--effective-yield relation 
                 derived by \citet{THK04} with data for individual 
                 gas-rich galaxies from the sample of \citet{KH05}. 
                 Large circles indicate the spiral galaxies in the 
                 sample of \citet{Ga02}. 
                 A filled symbol indicates that the 
                 gas fraction is taken from \citet{Ga02}. The
                 dot-dashed curve is the mass--yield relation for the
                 galactic wind model from \citet{THK04}.}
      \label{f:mstaryeffo}  
   \end{figure}

   We also compare the effective oxygen yields. Figure~\ref{f:mstaryeffo}
   shows that the mass--yield relation from \citet{THK04} is quite compatible
   with individual measurements, as was also shown by them by their
   choice of samples from the literature. A slight tendency to give higher
   values is also apparent in fig.~8 of \citet{THK04}. There is an appreciable
   dispersion for a given galactic mass, especially for masses below about
   $10^8 \msun$. But despite this dispersion, there is a strong tendency for
   small galaxies to have also low effective yields. Such a feature may be
   the consequence of a galactic wind scenario as proposed by \citet{THK04} as
   well as a signature for a variable IGIMF.
   One notes that most of the galaxies less massive than $10^9 \msun$
   have higher effective yields than predicted by the galactic wind
   model adopted by \citet{THK04}. The same holds true for our models,
   of course. Apart from larger uncertainties in the observational data
   expected from these dwarf galaxies, and the possibility that in 
   these objects the true gas content could have been overestimated
   due to the presence of \HI~envelopes \citep[cf. the example of Leo~A 
   discussed by][]{KH05}, our model offers a possible physical
   explanation for such a trend: bursts of intense star formation
   result in higher yields and higher metallicities than continuous
   star forming to the same present gas fraction. Nearby dwarf galaxies
   have quite bursty star formation histories \citep{Gr01}, which
   would tend to favour such an interpretation. Detailed studies
   of individual galaxies will be necessary to confirm this possibility.    

   \subsection{Scatter in the relations}

   There is a substantial spread about the mass--metallicity relation, and
   an even larger one about the mass--gas-fraction relation. It is not the aim 
   of this work to model this scatter, nor to provide physical explanations.
   But the chemical models permit to make the following statement: one could
   account for the spread in the gas fractions (Fig.~\ref{f:inf1mhifgas}) by 
   invoking, for instance, an intrinsic variation of a factor of 2 in the 
   star formation time scale for galaxies of the same mass, as indicated by
   the three series of models.
   The spread of the oxygen abundances from these models in
   Fig.~\ref{f:inf1moh} is somewhat larger than the dispersion found by
   \citet{THK04}. Because of the relationship $Z = y \ln(f_{\rm gas})$ between
   metallicity and fraction which exists in a closed-box model but is still
   closely valid in monotonic infall models, an increase in $\ln(f_{\rm gas})$,
   caused by a decrease in the SFR, gives rise to an increase in metallicity. 
   In our models with a variable IGIMF the lower SFR implies a smaller yield,
   thus the metallicity does not increase as much as if the yield were
   constant. While this can be seen in the results, it could constitute a
   possible test for the validity of our scenario: provided that gas 
   fractions and abundances can be measured with sufficient accuracy
   so as to determine genuine differences in individual objects, and
   if a correlation between deviations in abundances and gas fractions
   is found, our scenario would predict that the deviations in
   abundance should be less than in gas fractions. It is obvious that
   this quite subtle effect is not an easy test.  

\begin{figure}
      \centering
      \includegraphics[angle=-90,width=10cm]{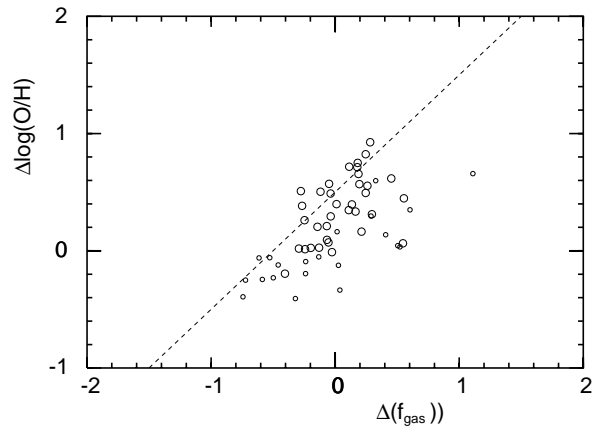}
      \caption[]{The deviations of the oxygen abundances from the mean
                 mass--metallicity relation of \cite{Ga02} as a function
                 of the deviations of the logarithmic gas fractions from
                 the assumed relation shown in Fig.~\ref{f:mhifgaso}. 
                 Large circles indicate the spiral galaxies in the 
                 sample of \citet{Ga02}. 
                 The dotted line indicates the slope of unity which would be
                 expected for closed-box chemical evolution with the same
                 constant effective yield.}
      \label{f:deltafz}  
   \end{figure}

   If one assumed a relation between galactic \HI~mass and the gas fraction in
   our sample of individual galaxies, the deviations from this relation are 
   correlated with the deviations of the oxygen abundances from the
   mass--metallicity relation of \citet{Ga02}, as shown in
   Fig.~\ref{f:deltafz}.
   This correlation is only marginally significant (correlation coefficient 0.6
   for all objects and 0.5 for the spiral galaxies); but as the dotted line
   indicates, the trend is in good agreement to what one should expect if all 
   galaxies had the same yield and evolved as nearly closed boxes.

\section{Conclusions}
   \label{se:conclu}

   The closed box model as well as the models with monotonically decreasing
   infall provide a remarkable match to the observed median mass--metallicity
   relation. This is in strong contrast to models with a constant IMF which
   all have the same effective yield for all galactic masses and all star
   formation time scales and therefore do not provide a reasonable match.
   In the case of the infall models also the effective yields match the 
   observed relation very reasonably. The match with the observations is
   independent of the adopted infall time scale.
   For models with a starburst the resulting IGIMF includes a higher number of
   massive stars. In comparison to continuous star formation the average metal
   yield will therefore be larger. This leads to to both higher oxygen
   abundances and higher effective yields in strong starbursts. It must be
   noted that the rather schematic model of starbursts used here mainly shows
   the tendency of the results for extreme burst-type star formation
   histories. Regardless of the details it is important that these
   results predict higher metallicities and higher oxygen yields for
   galaxies with strongly fluctuating star formation histories. 
   Our models -- closed box and infall -- thus constitute an upper
   limit of the effects due to a variation of the IGIMF on the SFR.
   By combining the effects of gas outflows and a variable IGIMF a steeper
   mass--metallicity relation would be achieved than the ones produced by
   either process alone. But in dwarf galaxies strong fluctuations in
   star formation would lead to higher yields thereby compensating
   this and flattening the stellar-mass--metallicity relation to the
   observed slope. 

   It may be possible to test our model in more detail provided that gas 
   fractions and abundances can be measured with sufficient accuracy
   so as to determine genuine differences in individual objects. If a
   correlation between deviations in abundances and gas fractions is
   found, our scenario would predict that the deviations in abundances
   should be less than in gas fractions.

   The idea of a SFR-dependent IGIMF provides a most attractive way to 
   explain the observed relations between metallicity and galactic mass
   as well as the one between effective yield and galactic mass. The
   main effect at work is the lowering of the effective upper mass
   limit of stars for low star-formation rates, which reduces the
   number of supernovae of type II.

   Our results demonstrate a good agreement with the mass-metallicity data,
   even for $\beta = 2$, without needing to invoke outflows. This may suggest
   that the importance of outflows for shaping the mass--metallicity relation
   may need reconsideration. Thus, the oxygen yield becomes a sensitive
   function of the SFR but also of the cluster MF power-law index
   $\beta$. Low-mass galaxies therefore appear chemically less-evolved than
   massive galaxies despite having the same ages.

   If one wanted to explain the mass--metallicity relation by an explicit
   variation of the slope of the IGIMF on galaxy mass, one would need
   to vary the exponent $\alpha$ above about 1.5 $M_{\odot}$ from 2.5 for
   galaxy masses of about $5\times10^8\,\msun$ to 2.3 for
   $2\times10^{11}\,\msun$, which is the range of the data covered by the
   observations of \citet{THK04}. These changes of the slope IGIMF $\alpha$
   correspond to the 'minimal' scenario proposed in \citet{WK05a}.

   In a recent paper \citet{LSC06} investigate the mass--metallicity
   relation with a more recent sample of dwarf galaxies with Spitzer data. The
   sample is in principle in agreement with the data we use and form an
   extension to lower galactic masses. They find a rather flat mass--yield
   relation noting it ''is difficult to explain, if galactic winds are
   ubiquitous in dwarf galaxies.'' Therewith they indicate that other
   solutions rather than galactic winds might still be possible. Here
   the flat stellar-mass--yield relation is explainable with a
   SFR-dependent IGIMF by invoking bursty star-formation histories for dwarfs.

   Furthermore in another recent paper \citep{Da06} it has been shown that
   neither outflows nor accretion of low-metallicity gas can reproduce the low
   effective yields observed in low mass galaxies. It is conceivable that a
   variable IGIMF could help to explain this problem.

   Our models indicate that massive spirals and ellipticals should
   show less variations in the IGIMF and thus metallicities due to their
   high average SFR. This is observed in nearby galaxies as is seen in
   fig.~9 in \citet{Ga04} or in the sample of 53000 galaxies plotted
   in fig.~6 in \citet{THK04}. The direct consequence of a dependence of the
   IGIMF on the SFR is a strong dependence of the oxygen abundance on the mass
   of a galaxy. It alone would already suffice to explain the observed
   mass--metallicity relation. Thus it would impose lesser demands on the loss
   of metals by outflows.

   To verify whether the SFR-dependence of the IGIMF could indeed be the 
   origin or at least part of the explanation of the mass--metallicity relation
   of galaxies, we are in the process of comparing model predictions with
   other observational properties, among them the abundances of other
   elements. The variation of the IGIMF affects strongly the massive
   stars but only weakly the intermediate mass stars. Thus we would
   expect that elements produced only in massive stars will vary as
   strongly as oxygen, but that elements with a strong contribution
   from intermediate mass stars -- such as carbon, nitrogen, and iron
   -- will vary to a lesser extend. This will be the subject of a
   future study.

  \section*{Acknowledgements}
    We thank Simone Recchi for discussions, and our anonymous referee
    for numerous detailed comments.
    This research has been supported by DFG grant KR1635/3 and the Chilean
    FONDECYT grant 3060096. 

  \bibliography{mybiblio}
\bsp
  \label{lastpage}

\end{document}